\documentclass{article}
\usepackage[preprint]{spconf}
\usepackage{amsmath,graphicx,multirow,hyperref,listings,bm}

\newcommand{\size}[2]{{\fontsize{#1}{0}\selectfont#2}}
\newcommand{\commentout}[1]{}
\newcommand{\argmax}[1]{\mathop{\arg\max}\limits_{#1}}

\title{Acoustic BPE for Speech Generation with Discrete Tokens}

\name{Feiyu Shen, Yiwei Guo, Chenpeng Du, Xie Chen, Kai Yu$^*$\thanks{*Corresponding author.}}
\address{MoE Key Lab of Artificial Intelligence, AI Institute\\
X-LANCE Lab, Department of Computer Science and Engineering\\
Shanghai Jiao Tong University, Shanghai, China\\
\texttt{\size{9}{
\{francis\_sfy,yiwei.guo,duchenpeng,chenxie95,kai.yu\}@sjtu.edu.cn
}}
}

\begin{document}
% default is 10pt
\ninept
\maketitle

% Abstract
% 100-150 words
\begin{abstract}

Discrete audio tokens derived from self-supervised learning models have gained widespread usage in speech generation. However, current practice of directly utilizing audio tokens poses challenges for sequence modeling due to the length of the token sequence. Additionally, this approach places the burden on the model to establish correlations between tokens, further complicating the modeling process. To address this issue, we propose {\em acoustic BPE} which encodes frequent audio token patterns by utilizing byte-pair encoding. Acoustic BPE effectively reduces the sequence length and leverages the prior morphological information present in token sequence, which alleviates the modeling challenges of token correlation. Through comprehensive investigations on a speech language model trained with acoustic BPE, we confirm the notable advantages it offers, including faster inference and improved syntax capturing capabilities. In addition, we propose a novel rescore method to select the optimal synthetic speech among multiple candidates generated by rich-diversity TTS system. Experiments prove that rescore selection aligns closely with human preference, which highlights acoustic BPE's potential to other speech generation tasks.

\end{abstract}

% keywords
\begin{keywords}
discrete audio token, byte-pair encoding, language modeling, rescore
\end{keywords}

% Introduction
\section{Introduction}
\label{sec:intro}
The emergence of self-supervised learning (SSL) models in the audio domain has introduced a new option for audio feature selection. For instance, wav2vec~\cite{schneider2019wav2vec} utilizes a contrastive loss objective to exploit high-level representation from audio signals. HuBERT~\cite{hsu2021hubert} utilizes a masked prediction objective where discrete targets derive from $k$-means~\cite{lloyd1982least} clustering. W2v-BERT~\cite{chung2021w2v} combines both contrastive learning and masked prediction in an end-to-end fashion. Aside from modeling speech content in the audio signal, WavLM~\cite{chen2022wavlm} encodes speaker-related information through a denoising objective, where the model is asked to predict the pseudo labels of original audio that is overlapped by a noisy or secondary utterance clip. vq-wav2vec~\cite{baevski2020vq} and wav2vec 2.0~\cite{larochelle2020wav2vec2} use vector quantization to learn a discrete representation of audio signals. 

Discretization further compresses redundant information and enables the direct application of algorithms from natural language processing (NLP) communities. With $k$-means clustering or vector quantization, these aforementioned SSL features can be utilized as pseudo text for speech generation purposes. GSLM~\cite{lakhotia2021generative} extracts SSL features from pretrained HuBERT which is then discretized using $k$-means clustering. These discrete tokens are utilized as inputs to Tacotron2~\cite{shen2018natural} for synthesizing mel-spectrograms. VQTTS~\cite{du2022vqtts,du2023multispeaker,du2023speaker} replaces the mel-spectrogram with vq-wav2vec tokens and a 3-dimensional prosody feature to bridge the acoustic model and the vocoder, achieving competitive synthesis quality compared to its mel-spectrogram counterpart. 
% Building upon VQTTS, TN-VQTTS~\cite{du2023speaker} further improves speaker adaptation by incorporating timbre-normalized vector quantized acoustic features. 
AudioLM~\cite{borsos2023audiolm} trains a language model on tokens discretized by w2v-BERT and $k$-means clustering, which demonstrates the effectiveness of discrete audio tokens as textual units in language modeling. This observation is further substantiated by SPEAR-TTS~\cite{Kharitonov2023speak}, which leverages discrete audio tokens as pseudo-textual units in low-resource scenarios.

However, the existing approaches directly utilize audio tokens, which poses challenges in sequence modeling. This is primarily due to the typically lengthy token sequences and the reliance on the model to capture correlations between tokens. While some solutions have been proposed to mitigate this issue, such as removing sequential repetitions of units used in \cite{lakhotia2021generative,borsos2023audiolm}, these approaches suffer from corruptive encoding, making them unsuitable for speech generation tasks. In text language modeling, this problem is relieved by combining a segment of consecutive tokens according to a certain rule. For example, 
% $n$-gram~\cite{chen1996empirical} merges $n$ consecutive characters into a single token, capturing token correlation within a fixed size window. 
Byte-pair encoding (BPE)~\cite{devlin2019bert} dynamically creates subword units based on frequency, encoding morphological information.

Taking inspiration from text-based approaches, we propose acoustic BPE for speech generation tasks, which extends BPE to discrete audio token sequences to reduce sequence length and leverage the morphological information present in token sequence. Previous studies have investigated the effectiveness of acoustic BPE in SSL model pretraining and automatic speech recognition (ASR). For instance, HuBERT-AP\cite{ren2022speech} employs BPE to encode the pseudo target label used in HuBERT pretraining to bridge the gap between audio signal and natural language. However, it does not extend BPE to the inference procedure. \cite{chang2023exploration} explores the improvements brought by acoustic BPE in the ASR task. \cite{hayashi2020discretalk} formulates text-to-speech (TTS) as a machine translation task, where the discrete VQVAE\cite{tjandra2019vqvae} sequence is encoded by BPE. However, less attention has been paid to exploring the detailed benefits brought by acoustic BPE in speech generation tasks.

In this study, we employ our proposed acoustic BPE to train a language model, which we refer to as the speech language model (SLM). The SLM serves as a generative model with various applications, including speech continuation and speech evaluation. Through comprehensive investigations, we uncover several notable benefits resulting from the use of acoustic BPE, including faster inference, better syntax capturing abilities, and improved diversity and richness in generation. These benefits can enhance various speech generation tasks, including text-to-speech, voice cloning, and speech enhancement. As an example, we introduce a novel rescore method designed for speech generation utilizing acoustic BPE. This method leverages the speech language model to evaluate the quality of different candidates generated by rich-diversity TTS systems. By selecting the optimal synthetic speech, our rescore method achieves a balance between diversity and naturalness. Experimental results demonstrate that the rescore selection closely aligns with human preference, further highlighting the effectiveness of acoustic BPE for other speech generation tasks.

In Section~\ref{sec:method}, we introduce the acoustic BPE and the speech language model used in experiments. We detailed our investigations on the SLM in Section~\ref{sec:experiment} and present the novel rescore method in Section~\ref{sec:exp_tts_rescore} as an application of acoustic BPE in speech generation tasks.

% Method
\section{Speech language model with acoustic BPE}
\label{sec:method}
In this section, we first give a comprehensive explanation of acoustic BPE (aBPE). Then, we introduce the unconditional speech language model (SLM) used in our experiments. Finally, we present the novel rescore method that effectively selects the optimal synthetic speech from various candidates generated by rich-diversity TTS.

\subsection{Acoustic BPE}
\label{sec:method_acoustic_bpe}
\begin{figure}[!h]
\centering
\includegraphics[width=0.8\linewidth]{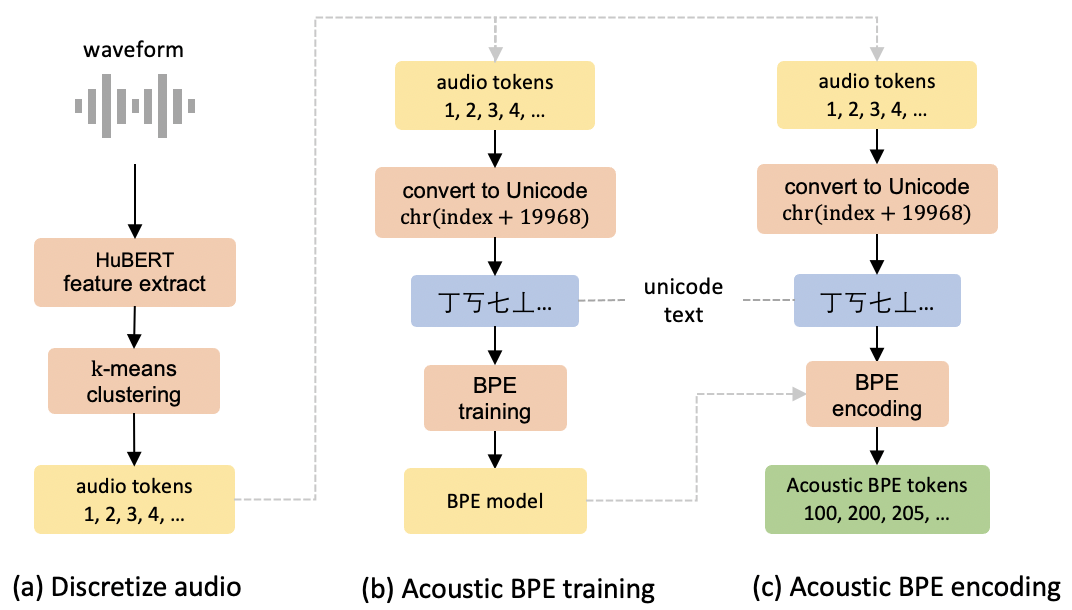}
\caption{Acoustic BPE training and encoding. (a) discretize audio into tokens leveraging the pretrained HuBERT model and $k$-means clustering. (b) convert discrete tokens into Unicode text for BPE training. (c) encode with the trained BPE model to obtain acoustic BPE tokens.}
\label{img:acoustic_bpe}
\end{figure}

On text corpus, the byte-pair encoding (BPE) works by initiating a vocabulary that contains all unique characters in the training text. Then it iteratively merges the most frequent character pair into one unit which is added to the vocabulary. It finishes till a desired vocabulary size is achieved. To adopt the BPE algorithm on discrete audio tokens, we first convert the token sequence into Unicode text, then apply BPE training and encoding on the Unicode text. The detailed process is depicted in Figure~\ref{img:acoustic_bpe} and involves the following steps:
\begin{enumerate}
    \item We discretize audio into tokens utilizing pretrained HuBERT models and $k$-means clustering.
    \item Then we convert the token sequence to Unicode text by mapping integer to common Chinese characters located within the Unicode region $\text{4E00}\sim\text{9FFF}$. This region contains $20992$ Chinese characters, which is sufficient for our purposes.
    \item Training BPE model on the obtained Unicode text with desired vocabulary size.
    \item Encode Unicode text to acoustic BPE tokens with the trained BPE model.
\end{enumerate}

\subsection{Speech language model}
\label{sec:method_slm}
The language model (parameterized by $\theta$) is designed to estimate the probability of a given sequence $\bm{x}=\{x_1,...,x_t\}$. It can be expressed as:
\begin{equation}
p(\bm{x}|\theta)=\prod_{i=1}^{t}p(x_i|x_{<i},\theta)
\end{equation}
Given a sequence, the language model can calculate the probability of this sequence or predict the probability distribution of the next token to perform tasks such as continuation. The speech language model (SLM) extends language modeling to discrete audio tokens. However, unlike previous approaches \cite{lakhotia2021generative, borsos2023audiolm, maiti2023speechlmscore} that directly utilize discrete audio tokens, our speech language model leverages acoustic BPE. This approach inherits the advantages of BPE encoding, such as enhanced sequence modeling and improved syntax capturing abilities. We use a decoder-only Transformer similar to \cite{lakhotia2021generativeborsos2023audiolm} and briefly explain the process of training and generating continuations.

\noindent \textbf{Training:} we first discretize audio into tokens utilizing HuBERT and $k$-means clustering, then, we encode these tokens into acoustic BPE tokens as in Section~\ref{sec:method_acoustic_bpe}. The SLM (parameterized by $\theta$) is optimized to maximize the joint probability of aBPE sequence $\bm x$:
\begin{equation}
\theta=\argmax{\theta} p(\bm{x}|\theta)=\argmax{\theta}\prod_{i=1}^{t}p(x_i|x_{<i},\theta)
\end{equation}

\noindent \textbf{Generating continuation:}
The speech language model can generate continuations after a given prompt $\bm p=\{p_1,..,p_l\}$ by autoregressively sampling the next token $x_{i}$:
\begin{equation}
x_{i} \sim p(x_i|x_{<i},\bm{p},\theta)
\end{equation}

\subsection{Acoustic rescoring with SLM}
\label{sec:method_acoustic_rescore}
Recent TTS systems can generate diverse synthetic speeches. However, the naturalness of these synthetic speeches can vary considerably. To strike a balance between diversity and naturalness, one approach is to generate multiple candidates from the TTS system and manually select the most natural one. Nonetheless, this manual evaluation process is laborious and time-consuming. To address this issue, we propose a novel rescore method leveraging the SLM to select the optimal synthetic speech that ensures both diversity and naturalness.

With multiple synthetic speeches $\bm{u}^1,...,\bm{u}^n$, we perform rescoring with SLM by first calculating the probability $y^i$ of each speech $\bm{u}^i$ and then select the synthetic speech with the highest probability as the optimal one as follows:
\begin{align}
\bm{x}^i=\{x^i_1,...,x^i_t\}=\text{Tokenizer}(\bm{u}^i)\\
y^i=p(\bm{x}^i|\theta)=\prod_{k=1}^{t}p(x^i_k|x^i_{<k},\theta)\\
\bm{u}^{best}=\bm{u}^{\argmax{i} y^i}
\end{align}
\noindent where the Tokenizer first discretizes audio into tokens by HuBERT and $k$-means, and then encodes tokens by acoustic BPE.

% no room for this image
\commentout{
% rescore image
\begin{figure}[!h]
\centering
\includegraphics[width=0.8\linewidth]{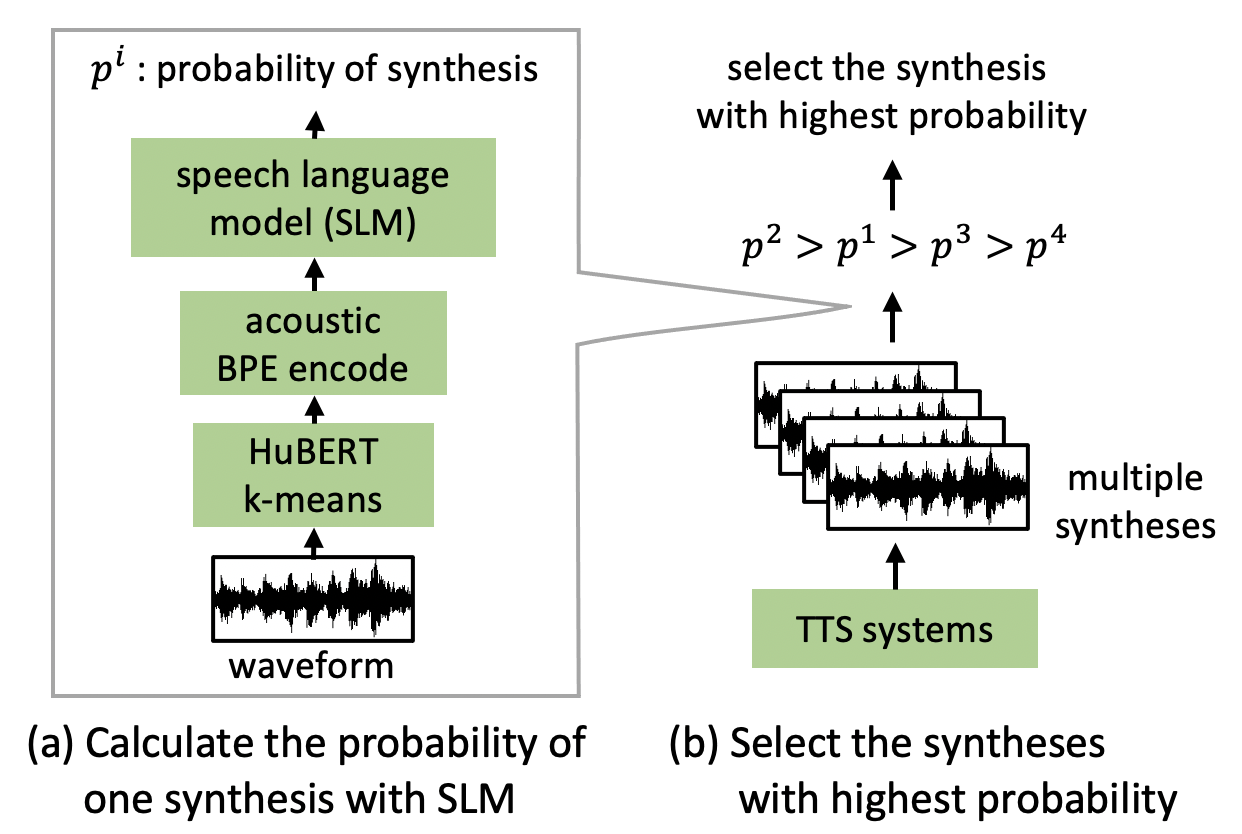}
\caption{We rescore multiple audio samples from TTS by: (a) tokenizing them into acoustic BPE tokens and calculating probability with SLM. (b) We select the one with the highest probability as the optimal synthetic speech.}
\label{img:tts_rescore}
\end{figure}
}

% Experiment
\section{Experiment}
\label{sec:experiment}
% setup
% ----------------------------------------------
\subsection{Experimental setup}
\textbf{Dataset:} We conduct experiments using two datasets: LibriTTS~\cite{zen2019libritts}, which consists of 580 hours of speech data, and the 6k-hour clean subset of LibriLight~\cite{kahn2020librilight}. For LibriLight, we utilize the official scripts to segment the original long recordings into utterances of 60 seconds. 
All the speech waveforms are downsampled to 16kHz before training and inference. 

\noindent \textbf{Acoustic BPE:} We extract speech features from the final layer of pretrained HuBERT Large model\footnote{\url{https://github.com/facebookresearch/fairseq/blob/main/examples/hubert}}. Then we train $k$-means clustering with 2000 centroids on a randomly selected 100-hour subset of speech features from the LibriTTS train-960 subset. 
All the speech data from LibriTTS and LibriLight is encoded to discrete tokens with the trained $k$-means model as illustrated in Figure \ref{img:acoustic_bpe} (a). 

The acoustic BPE model is trained on the LibriTTS train-960 subset. We first convert discrete tokens to Unicode text and train the acoustic BPE model using SentencePiece\footnote{\url{https://github.com/google/sentencepiece}} with the desired vocabulary size as outlined in Section \ref{sec:method_acoustic_bpe}. 
Then we use the trained BPE model to encode the discrete tokens of LibriLight into acoustic BPE tokens. 
In the following experiments, we compare four acoustic BPE variants: without acoustic BPE and acoustic BPE of vocabulary size 5k, 10k, and 20k. 

\noindent \textbf{Speech language model:} We use a decoder-only Transformer with 12 layers, 16 attention heads, an embedding dimension of 1024, and a T5-style~\cite{raffel2020exploring} relative positional embedding mechanism. 
During training, we use random cropping to equivalent input lengths of 15 seconds. 
Models with four acoustic BPE variants are trained on LibriLight 6k subset for 10 epochs with learning rate linearly increases from 0 to $1\times 10^{-5}$ for the first epoch and cosine decay to $1\times 10^{-6}$ for subsequent epochs. 

\noindent \textbf{Decoding discrete tokens to waveform:} We train a CTX-vec2wav vocoder proposed in \cite{du2023unicats} to decode waveform from discrete tokens. The CTX-vec2wav is configured to have a frameshift of 20ms and kernel sizes (16,10,8,4) in its HifiGAN upsampling layers. It is trained on LibriTTS train-960 subset with discrete speech tokens discretized by HuBERT and $k$-means clustering. When decoding, acoustic BPE tokens are first decoded and then synthesized to waveform. In all experiments,  we use the same speaker (speaker ID 121) as a prompt during vocoding.

In the subsequent subsections, we compare of the inference speed of speech continuation using various acoustic BPE variants in subsection~\ref{sec:exp_sequence_compression}. Following that, we examine the syntax capturing abilities of the SLM in subsection~\ref{sec:exp_improved_syntax_modeling} and generation diversity and richness in subsection~\ref{sec:exp_improved_syntax_modeling} and \ref{sec:exp_information_entropy}, respectively. Lastly, in subsection~\ref{sec:exp_tts_rescore} we demonstrate the effectiveness of our novel rescore method, emphasizing the advantages conferred by the utilization of acoustic BPE.
\subsection{Inference speedup}
\label{sec:exp_sequence_compression}
Acoustic BPE combines frequent tokens into one single unit, significantly reducing the sequence length. In Table~\ref{tab:sequence_compression_wo_text}, we present a comparison of the average sequence length for LibriLight-6k across four acoustic BPE variants: without acoustic BPE and acoustic BPE with vocabulary size 5k, 10k, and 20k. Also, We assess their inference speed of speech continuation with the SLM. We use the first 3 seconds of a randomly selected utterance from the LibriTTS test-clean subset as a prompt and generate 10 continuations each of 20 seconds long on an NVIDIA V100 GPU with 32GB memory. We compare the speedup brought by acoustic BPE in Table~\ref{tab:sequence_compression_wo_text}.

\begin{table}[!h]
\centering
% \setlength{\tabcolsep}{4.5pt}
% \renewcommand{\arraystretch}{1.1}
% \resizebox{1\linewidth}{!}{
\begin{tabular}{cccc}
\hline
\textbf{Encoding} & \textbf{\begin{tabular}[c]{@{}c@{}}Num. \\ aBPE\end{tabular}} & \textbf{\begin{tabular}[c]{@{}c@{}}Avg. \\ sequence length\end{tabular}} & \textbf{\begin{tabular}[c]{@{}c@{}}Inference \\ speedup\end{tabular}} \\ \hline \hline
w/o aBPE & - & 2513.8 & 1.0x \\ \hline
\multirow{3}{*}{aBPE} & 5k & 1547.0 & 2.8x \\
 & 10k & 1241.0 & 3.8x \\
 & \textbf{20k} & \textbf{1053.0} & \textbf{5.0x} \\ \hline
\end{tabular}
% }
\caption{Comparison of sequence lengths and inference speedup across four acoustic BPE
 variants.}
\label{tab:sequence_compression_wo_text}
\end{table}

From Table~\ref{tab:sequence_compression_wo_text}, the employment of acoustic BPE compresses the sequence by 1.6 to 2.4 times, which is believed to ease the sequence modeling. A shorter sequence also accelerates the autoregressive inference procedure by 2.8 to 5.0 times.

\subsection{Syntax capturing with acoustic BPE}
\label{sec:exp_improved_syntax_modeling}
Following GLSM~\cite{lakhotia2021generative} and AudioLM~\cite{borsos2023audiolm}, we assess the syntax modeling capability of SLM by distinguishing between a pair of syntactically correct and non-correct utterances. Similar to the aforementioned rescore method, we consider the utterance with higher probability as syntactically correct. To construct such test pairs, we filter out too short utterances from the LibriTTS test-all subset and use the remaining 5497 utterances as syntactically correct ones. To create non-meaningful utterances, we randomly shuffle the words in the syntactically correct text.
% as depicted in Figure~\ref{img:lm_test_case}. 
Both syntactically correct and non-correct utterances are synthesized to waveforms using VQTTS~\cite{du2022vqtts} trained on LibriTTS train-960. Subsequently, the SLM is used to classify the two utterances in each test case into syntactically correct and non-correct. This classification accuracy is referred to as syntax accuracy. The syntax accuracy among four acoustic BPE variants is presented in Table~\ref{tab:lmacc_vert}.

\commentout{
\begin{figure}[!h]
    \centering
    \includegraphics[width=\linewidth]{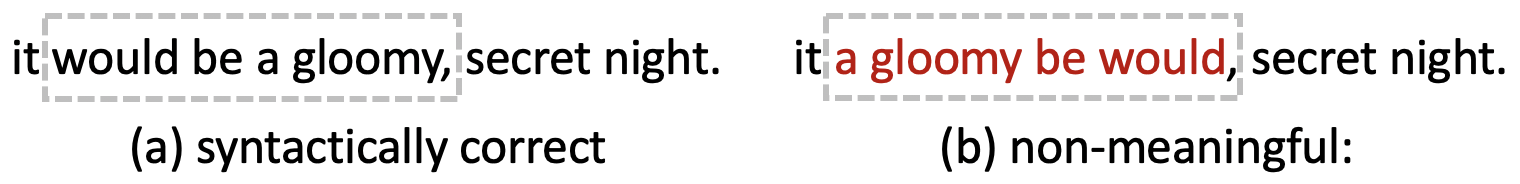}
    \caption{Example for constructing syntactically non-correct utterances.}
    \label{img:lm_test_case}
\end{figure}
}

Moreover, we examine the diversity of generated speech content similar to \cite{lakhotia2021generative}. We randomly select 3 utterances from the LibriTTS test-clean subset and use the first 3 seconds of each utterance as a prompt to generate a 20-second continuation using the SLM. For each prompt, we repeat the continuation process 50 times, resulting in a total of 150 utterances. We use the open-sourced whisper\footnote{\url{https://github.com/openai/whisper/blob/main}} to transcribe speech into text. To evaluate the diversity of these texts, we employ the $n$-gram VERT metric proposed in \cite{lakhotia2021generative}. This metric is a geometric mean of the $n$-gram self-BLEU and $n$-gram auto-BLEU scores, measuring the $n$-gram diversity both across sentences and within sentences. Higher VERT values indicate lower diversity. The results of $3$-gram VERT are presented in Table~\ref{tab:lmacc_vert}.

\begin{table}[!h]
\centering
\setlength{\tabcolsep}{4pt}
% \renewcommand{\arraystretch}{1.1}
% \resizebox{1\linewidth}{!}{
\begin{tabular}{cccc}
\hline
\textbf{Encoding} & \textbf{Num. aBPE} & \textbf{Syntax acc.($\uparrow$)} & \textbf{3-gram VERT($\downarrow$)} \\ \hline \hline
w/o aBPE & - & 75.35\% & 6.99 \\ \hline
\multirow{3}{*}{\textbf{aBPE}} & 5k & 83.45\% & 7.55 \\
 & \textbf{10k} & \textbf{85.37\%} & 5.97 \\
 & 20k & 85.34\% & \textbf{4.71} \\ \hline
\end{tabular}
% }
\caption{Comparison of syntax probing accuracy and generation diversity among four acoustic BPE(aBPE) variants.}
\label{tab:lmacc_vert}
\end{table}

As expected, the incorporation of acoustic BPE enhances the speech language model's ability to accurately capture syntax structure and effectively model a wider range of diverse syntax patterns. 
\subsection{Generation richness with acoustic BPE}
\label{sec:exp_information_entropy}
With acoustic BPE, the SLM can generate diverse outputs that are different and vary in content. In this section, we explore other enhancements brought by acoustic BPE, specifically in generating outputs that can convey more information within a limited time. 

To quantify the informativeness of the SLM, we employ cross-entropy $H$ of the text content from the SLM generated speech with respect to a well trained text language model. The text LM can be viewed as an approximation of the true distribution of all meaningful text-content. Hence, regarding each SLM with different aBPE as an information source and assuming speech recognition does not introduce much errors, we can view cross-entropy as the measurement of the information contained in the synthetic speech. The cross-entropy is calculated as follows: Firstly, we generate a set of prompted continuations $\{\bm{u}_1,...,\bm{u}_n\}$ from the SLM (parameterized by $\theta$), which are transcribed into text with ASR. Next, for each continuation, we calculate the log-probability of its transcript with a pretrained text language model (parameterized by $\gamma$). The cross-entropy $H(\text{SLM}_{\theta}|\text{TextLM}_{\gamma})$ is obtained by averaging the negative log-probability over all generated continuations.

\begin{align}
\{\bm{u}_1,...,\bm{u}_n\} &\sim p(\bm{u}|\theta)\\
H(\text{SLM}_{\theta}|\text{TextLM}_{\gamma}) = - \frac{1}{n} & \sum_{i=1}^{n} \log p(\text{ASR}(\bm{u}_i)|\gamma)
\end{align}
In our experiments, we crop the first 3 seconds from a randomly select utterance in the LibriTTS test-clean subset as prompt and generate 150 continuations each lasting 20 seconds. we use Whisper to transcribe speech to text and use a pretrained text LM\footnote{\url{https://github.com/pytorch/fairseq/tree/master/examples/language_model}} to calculate cross-entropy. We present the results in Table~\ref{tab:info_entropy}.

\begin{table}[!h]
\centering
% \setlength{\tabcolsep}{3pt}
% \renewcommand{\arraystretch}{1.05}
% \resizebox{1\linewidth}{!}{
\begin{tabular}{ccc}
\hline
\textbf{Encoding} & \textbf{Num. aBPE} & \textbf{cross-entropy} \\ \hline \hline
w/o aBPE & - & 352.4 \\ \hline
\multirow{3}{*}{\textbf{aBPE}} & 5k & 409.1 \\
 & 10k & 457.6 \\
 & \textbf{20k} & \textbf{469.6} \\ \hline
\end{tabular}
% }
\caption{Comparison of cross-entropy among four acoustic BPE(aBPE) variants.}
\label{tab:info_entropy}
\end{table}

The usage of acoustic BPE increases the amount of information conveyed by the SLM within a limited time, indicating that acoustic BPE contributes to the richness of generation.

\subsection{Text-to-speech rescoring with SLM}
\label{sec:exp_tts_rescore}
% valle checkpoint:
% \footnote{\url{https://drive.google.com/file/d/1pKvS56NnzVCYqhbjoFqL_JQ8i32UixbL/view?usp=sharing}}

The benefits discussed above regarding acoustic BPE highlight its potential application in speech generation tasks. Here, we introduce a novel rescore method that selects the optimal synthetic speech from many candidates generated by diverse TTS systems. This method aims to balance between maintaining diversity and ensuring naturalness in the generated speech.

We use a rich-diversity TTS system\footnote{\url{https://github.com/lifeiteng/vall-e}} trained on LibriTTS train-960 subset. We randomly select 106 utterances from the LibriTTS test-clean subset and synthesize each utterance 5 times. Before proceeding to rescore with SLM, we conduct a subjective listening test where 10 listeners are asked to rank the 5 synthetic speeches based on naturalness. Subsequently, We use SLM to rescore the 5 synthetic speeches as described in Section~\ref{sec:method_acoustic_rescore}. To quantitatively evaluate how close rescore selection aligns with human preference, we calculate the top-$x$ accuracy which denotes the success rate of rescore selection appearing within the top-$x$ of human rankings. We compare the top-$1$ to top-$3$ accuracy with four acoustic BPE variants in Table~\ref{tab:rescore_acc}.

\begin{table}[!ht]
\centering
\setlength{\tabcolsep}{4pt}
% \renewcommand{\arraystretch}{1.1}
% \resizebox{1\linewidth}{!}{
\begin{tabular}{ccccc}
\hline
\textbf{Encoding} & \textbf{Num. aBPE} & \textbf{Top1 acc.} & \textbf{Top2 acc.} & \textbf{Top3 acc.} \\ \hline \hline
random & - & 20.0\% & 40.0\% & 60.0\% \\ \hline
w/o aBPE & - & 29.3\% & 52.9\% & 74.3\% \\ \hline
\multirow{3}{*}{\textbf{aBPE}} & 5k & 26.4\% & 50.7\% & 73.6\% \\
 & \textbf{10k} & \textbf{31.4\%} & \textbf{57.9\%} & \textbf{77.9\%} \\
 & 20k & 29.3\% & 55.0\% & 76.4\% \\ \hline
\end{tabular}
% }
\caption{Rescore accuracy among various acoustic BPE(aBPE).}
\label{tab:rescore_acc}
\end{table}

\begin{figure}[!h]
\centering
\includegraphics[width=0.7\linewidth]{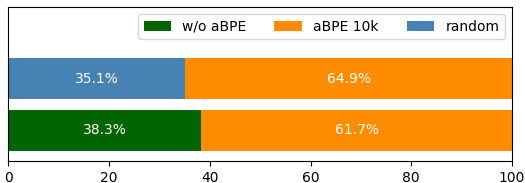}
\caption{Preference test. Above: random vs. aBPE 10k; Below: w/o aBPE vs. aBPE 10k.}
\label{img:pref_mos}
\end{figure}

As shown above, the rescore selection aligns with human preference. The use of acoustic BPE further improves rescore performance. Moreover, we conducted two preference tests to verify the effectiveness of rescore and acoustic BPE. The first preference test involves a random selection and the rescore selection from SLM with acoustic BPE 10k. The second test comprises selections from two acoustic BPE variants: without acoustic BPE and acoustic BPE 10k. Results are shown in Figure~\ref{img:pref_mos}.

The preference test further confirms the effectiveness of the rescore method and highlights the potential of acoustic BPE for speech generation tasks. 
% We believe that the improvement can be attributed to two factors. Firstly, the utilization of acoustic BPE reduces sequence length, thereby alleviating the modeling burden of SLM and helps with the limited attention span as mentioned in \cite{lakhotia2021generative}. Secondly, acoustic BPE enhances the syntax modeling capability. 

% no room for experiments on WavLM...

% Conclusion
\section{Conclusion}
\label{sec:conclusion}
In this work, we introduce the acoustic BPE to speech generation tasks.  We trained a generative speech language model using acoustic BPE and conducted thorough investigations on its property. These investigations uncover significant advantages associated with acoustic BPE. Firstly, the use of acoustic BPE results in shorter sequences, facilitating sequence modeling and accelerating autoregressive inference. Secondly, by leveraging the morphological information present in the token sequence, it alleviates the burden on the speech language model to construct token correlations, thereby enhancing its syntax capturing abilities and generation diversity and richness. Moreover, we demonstrate the application of acoustic BPE on a novel TTS rescore method, which selects the optimal one from multiple diverse TTS syntheses to strike a balance between diversity and naturalness. Experimental results provide empirical evidence of the effectiveness of acoustic BPE. These findings open up possibilities for applying acoustic BPE to other speech generation tasks such as text-to-speech, which can be explored in future research.

% Acknowledgements
\section{Acknowledgements}
This work was supported by China NSFC Project (No. 92370206), Shanghai Municipal Science and Technology Major Project \\(2021SHZDZX0102) and the Key Research and Development Program of Jiangsu Province, China (No.BE2022059).

% Reference
\vfill\pagebreak
\bibliographystyle{IEEEbib}
\bibliography{refs}

\end{document}